\documentclass[twocolumn,aps]{revtex4}
\usepackage{graphicx}

\usepackage{amssymb}
\usepackage[latin1]{inputenc}
\usepackage{bm}% bold math

\begin{document}

\title{Pseudo-diffusive magnetotransport in graphene}
\author{Elsa Prada$^{1}$, Pablo San-Jose$^{1}$, Bernhard Wunsch$^{2,3}$, and Francisco Guinea$^{3}$}

 %\email{elsa@tfp.uni-karlsruhe.de}

 \affiliation{$^1$Institut f\"ur Theoretische Festk\"operphysik and DFG-Center for Functional Nanostructures (CFN), Universit\"at Karlsruhe, D-76128 Karlsruhe, Germany\\
$^2$Departamento de F\'isica de Materiales, Universidad Complutense de Madrid, E-28040 Madrid, Spain.\\
$^3$Instituto de Ciencia de Materiales de Madrid, CSIC, Cantoblanco, E-28049 Madrid, Spain.}

\date{\today}

\begin{abstract}
Transport properties through wide and short ballistic graphene
junctions are studied in the presence of arbitrary dopings and
magnetic fields. No dependence on the magnetic field is observed at
the Dirac point for any current cumulant, just as in a classical
diffusive system, both in normal-graphene-normal and
normal-graphene-superconductor junctions. This pseudo-diffusive
regime is however extremely fragile respect to doping at finite
fields. We identify the crossovers to a field-suppressed and a
normal ballistic transport regime in the magnetic field - doping
parameter space, and provide a physical interpretation of the phase
diagram. Remarkably, pseudo-diffusive transport is recovered away
from the Dirac point in resonance with Landau levels at high
magnetic fields.
\end{abstract}

\maketitle

Low energy excitations in a monolayer of carbon atoms arranged in a
honeycomb lattice, known as a graphene sheet, have the remarkable
peculiarity of being governed by the 2D massless Dirac equation,
which is responsible for a variety of exotic transport properties as
compared to ordinary metals. Particularly striking is that for
clean, undoped graphene the density of states is zero, but not so
the conductivity, which remains of the order of the quantum unit
$e^2/h$ \cite{Netal04,Netal05b}. Another intriguing fact is that a
wide and short strip of undoped graphene exhibits
``pseudo-diffusive" transport properties in the absence of
electron-electron interactions and impurity scattering
\cite{Tworzydlo06}. By pseudo-diffusive it is meant that transport
properties are indistinguishable from those of a classical diffusive
system. These include the full transport statistics (in particular
the Fano factor $F=1/3$ and the conductance $G\propto W/L$
\cite{Tworzydlo06}, where $W$ is the width and $L$ is the length of
the graphene strip), the critical current \cite{Titov06} and I-V
characteristics \cite{Cuevas06} in Josephson structures, as well as
the relation of the normal metal-superconductor conductance to the
normal transmissions \cite{Akhmerov06}. The same behaviour can be
expected in bilayer graphene \cite{SB06}. In fact, all of the above
similarities can be explained by noting that at the Dirac point
(i.e., for undoped graphene) transport occurs entirely via
evanescent modes with a transmission that is equal to the diffusive
transport theory result (evaluated at $k_F l=1$, $l\equiv$ mean free
path \footnote{It is noteworthy that the diffusive transport theory
actually assumes weak disorder $k_F l\gg 1$. Hence Eq. (\ref{TRM})
is an extrapolation of the theory to a dirty metal limit.}) without
quantum corrections \cite{Tworzydlo06,Akhmerov06,Beenakker97},
\begin{equation}
T_{k_y}=\frac{1}{\cosh^2{k_y L}},\label{TRM}
\end{equation}
Here $k_y$ is the transverse momentum of the channel. In diffusive
systems the above relation holds independently of an externally
applied magnetic field in the limit of many channels (classical
limit), for which any quantum weak localization correction is
negligible \cite{Beenakker97}.

The question we raise here is: Does the diffusive behaviour of
ballistic graphene persist in the presence of a magnetic field? We
will show that for zero doping the equivalence is preserved for any
magnetic field. Remarkably for a ballistic system, the applied
magnetic field does not affect the transport statistics (for any
current cumulant) at the Dirac point. For graphene with disorder
this has recently also been demonstrated at the conductivity level
\cite{Gusynin05,Ostrovsky06,VGCE06}. At sufficiently strong magnetic
fields an exponentially small chemical potential is enough to enter
a field-suppressed transport regime. However, at resonance with the
Landau levels (LLs) the pseudo-diffusive behaviour is recovered for
all current cumulants. For even higher dopings one observes a final
crossover to the ballistic magnetotransport regime, since clean
graphene then resembles a ballistic normal metal.

The magnetic field introduces a fundamental quantum-mechanical
length scale known as the magnetic length $l_B=(\hbar/|e
\vec{B}|)^{1/2}$. In complete absence of scattering, localized LLs
are well formed and ballistic transport is suppressed. The only
contribution to transport in this regime comes from resonant
tunneling exactly at Landau energies. In usual metals, this happens
when the cyclotron diameter $2r_c=2l_B^2 k_F$ is smaller than the
relevant scattering length (set by system size, disorder or
temperature). If $2r_c$ is of the order of or larger than the
scattering length, then delocalized states contribute to transport
leading to Shubnikov--de Haas oscillations and the Quantum-Hall
effect. For wide and short ballistic strips the relevant scattering
scale is the strip length $L$, and scattering on lateral boundaries
can be ignored. In graphene, the lowest LL lies precisely at the
Dirac point. Besides, at this point $k_F=0$ and thus $r_c=0$
independently of the magnetic field. Therefore, in contrast to the
normal metal strip and to the high doping limit, at the Dirac point
no delocalized bulk transport should take place for any magnetic
field and resonant tunneling should be field independent.

To confirm this hand-waving picture we analyze theoretically
magnetotransport effects through normal-graphene-normal (N) and
normal-graphene-superconductor (NS) wide ballistic junctions at
arbitrary dopings and magnetic fields. From an experimental point of
view, transport properties of lightly doped graphene in contact with
superconductors is currently being investigated \cite{Hetal06}.
Moreover, the properties of graphene in strong magnetic fields are
also a subject of great interest \cite{Novoselov05,Zhang05,PGN06},
in particular in relation to weak (anti)localization
\cite{Metal06,MG06,Metal06b,WLSBH06,NM06}. Here we consider a clean
graphene sheet of width $W$ (assumed to be the largest lengthscale
in the system) in the $y$-direction through which transport occurs
in the $x$-direction. For $x<-L/2$ it is covered by a normal
contact, and for $x>L/2$ it is covered either by a superconducting
contact or a normal one. The central region is lightly doped,
leading to a finite Fermi energy $\mu$ measured relative to the
Dirac point, which can be varied by an external gate voltage. The
contact regions are modeled as heavily doped graphene, with Fermi
energy $\mu_c$ conveniently fixed to infinity with respect to both
$\mu$ and the superconducting gap $\Delta$. The boundary conditions
in the $y$-direction are irrelevant \cite{Tworzydlo06} for large
aspect ratios $W/L\gg 1$. We choose periodic boundaries for
simplicity.
%In
%the NS case we assume for simplicity a step-like pair potential
%\footnote{To avoid intervalley scattering complications, the
%magnetic field jump should occur at length scales much bigger than
%the lattice spacing.}.
A constant external magnetic field $B$ is applied perpendicular to
the graphene sheet. We assume the electrodes to be magnetically
shielded, e.g. by covering them with materials with high-magnetic
permeability. In the Landau gauge we can write the vector potential
as $\vec{A}=(0,Bx,0)$ for $|x|<L/2$, and constant in the contact
regions. This gauge is convenient since the motions in x- and
y-direction are uncoupled and $k_y$ remains a good quantum number.
We neglect Zeeman splitting, so that the electron spin only enters
as a degeneracy factor of 2 in the following calculation. Finally,
we note that edge currents generally give a negligible contribution
to transport in the $W\gg L$ limit.

We will compute the N and the NS (Andreev) inverse longitudinal
resistivity, $\rho_{xx}^{-1}=G(L/W)$, expressed in terms of the
conductances \cite{Tworzydlo06,Akhmerov06}
\begin{equation}
G_\mathrm{N}=\frac{4e^2}{h}\sum_{k_y}T_{k_y},\;G_\mathrm{NS}=\frac{8e^2}{h}\sum_{k_y}\frac{T^2_{k_y}}{(2-T_{k_y})^2},\label{conductances}
\end{equation}
and the shot noise using corresponding expressions in terms of the
transmission for normal conducting contacts $T_{k_y}$
\cite{Blanter00}. Note that the above expressions for the NS case
are valid only if $T_{k_y}$ is left-right symmetric, which is not in
general true in the presence of a magnetic field. In our particular
setup it does indeed turn out to be symmetric. The transmission
through the central region is obtained by imposing current
conservation at the interfaces, which translates into continuity of
the wavefunction. In the chosen gauge the scattering problem is
effectively one-dimensional, the transverse mode profile $e^{ik_y
y}$ being the same in all regions, so we will only discuss the
x-dependence of the wavefunctions from now on.

The contact region eigenstates at energy $\epsilon=\hbar
v_F\sqrt{k_x^2+k_y^2}-\mu_c+\mu$ with respect
to the central strip Dirac point are given by %$\Psi^N_{k_x k_y
%s}(x)=(s\, z_{\mathbf{k}}^{-s},1)^T e^{ik_x x}$
\begin{equation}
\Psi^N_{k_x k_y s}(x)=\left(
\begin{array}{c} s\, z_{\mathbf{k}}^{-s}\\1\end{array}
\right)e^{ik_x x}.
\end{equation}
The spinor lives in the space of the two triangular sublattices that
conform the graphene hexagonal lattice, $s=\pm 1$ is the `valley'
quantum number (for the degenerate $K$ and $K'$ points), and
$z_{\mathbf{k}}\equiv\exp[i\arg (k_x+i k_y)]$, which tends to
$z_{\mathbf{k}}\approx\mathrm{Sign}(k_x)$ when $\mu_c\rightarrow
\infty$.

The spinor $\Psi^G_{\epsilon k_y s}(x)=[\phi^A_{\epsilon k_y
s}(x),\phi^B_{\epsilon k_y s}(x)]^T$ for the central region is
determined by the 1D Dirac equation
\begin{equation}
\left(\begin{array}{cc} 0&-i\hat a\\i\hat a^+&0
\end{array}\right)\left(\begin{array}{c} \phi^A_{\epsilon k_y s}(x)\\\phi^B_{\epsilon k_y s}(x)\end{array}
\right)= \lambda\sqrt{n_\epsilon} \left(\begin{array}{c}
\phi^A_{\epsilon k_y s}(x)\\\phi^B_{\epsilon k_y s}(x)\end{array}
\right),\label{Dirac}
\end{equation}
with $\hat a\equiv(\tilde x +\partial_{\tilde{x}})/\sqrt{2}$, $\hat
a^+\equiv(\tilde x -\partial_{\tilde{x}})/\sqrt{2}$, $\tilde x\equiv
x/l_B+k_y l_B$, $\lambda=\mathrm{sign}~\epsilon$ and
$n_\epsilon=(l_B|\epsilon|/\hbar v_F)^2/2$. Canonical relations
$[\hat a,\hat a^+]=1$ are satisfied. The above equation corresponds
to the $K$ valley ($s=1$), while the $s=-1$ equation is obtained by
swapping $\hat a$ and $\hat a^+$. Since the central region is
bounded, no integrability condition must be met and the
eigenspectrum of (\ref{Dirac}) is continuous. The usual LL
solutions, which correspond to integer $n_\epsilon$, are thus
complemented by a larger family of divergent wavefunctions,
typically localized around the interfaces $x=\pm
L/2$, with arbitrary $n_\epsilon\ge 0$ \cite{Martino}. %%($\epsilon$ is
%%arbitrary and given by the states $\Psi^N_{k_x k_y s}$ coming from
%%the contacts).
%For $n_\epsilon>0$ the general solutions to (\ref{Dirac}) can be
%expressed by convenient combinations $(h^{e
%(o)}_{n_\epsilon},ih^{o(e)}_{n_\epsilon\pm 1})^T$ of even and odd
%(in $\tilde x$) solutions $h^{e,o}_n(\tilde x)$ of the Klein-Gordon
%equation $a^+a h^{e(o)}_n(\tilde x)=nh^{e(o)}_n(\tilde x)$ [the
%square of (\ref{Dirac})], normalized so that $\hat a^+\,h^{e
%(o)}_n=\sqrt{n+1}\;h^{o(e)}_{n+1}$, $\hat a\,
%h^{e(o)}_n=\sqrt{n}\;h^{o(e)}_{n-1}$. It is possible to write closed
%form expressions for $h^{e,o}_n(\tilde x)$ with real $n> 0$ in terms
%of the confluent hypergeometric function
%${}_1F_1(a,b,z)=1+\frac{a}{b}\frac{z}{1!}+\frac{a(a+1)}{b(b+1)}\frac{z^2}{2!}+\dots$,
%and $S_n=\mathrm{Sign}\{\sin[\pi(n+1/2)/2]\}$:
%\begin{eqnarray*}
%h^e_n(\tilde x)&=&\sqrt{\frac{(n-1)!!}{\sqrt{\pi}n!!}}S_{n+1}\,e^{-\tilde x^2/2}\,{}_1F_1\left(-\frac{n}{2},\frac{1}{2},\tilde x^2\right),\\
%h^o_n(\tilde x)&=&\sqrt{\frac{2 n!!}{\sqrt{\pi}(n-1)!!}}S_n \,\tilde
%x\, e^{-\tilde
%x^2/2}\,{}_1F_1\left(-\frac{n-1}{2},\frac{3}{2},\tilde x^2\right).
%\end{eqnarray*}
%The $\epsilon=0=n_\epsilon$ case must be treated separately, giving
%wavefunctions in terms of $\exp[\pm\tilde{x}^2/2]$ instead.
At the Dirac point ($n_\epsilon=0$) the components of the spinor are
uncoupled, and the two eigenstates for $s=1$ are $\Psi^{G,1}_{0 k_y
1}(x)=[0,\exp(-\tilde{x}^2/2)]^T$ and $\Psi^{G,2}_{0 k_y
1}(x)=[\exp(+\tilde{x}^2/2),0]^T$. The $s=-1$ solution has
interchanged spinor components. At finite energy ($n_\epsilon>0$)
the solutions to Eq.~(\ref{Dirac}) become
\begin{eqnarray*}
\Psi^{G,1(2)}_{\epsilon k_y 1}(x)=\left(\begin{array}{c}\lambda
h^{e(o)}_{n_\epsilon-1}(\tilde{x})\\ih^{o(e)}_{n_\epsilon}(\tilde{x})\end{array}\right);
\Psi^{G,1(2)}_{\epsilon k_y -1}(x)=\left(\begin{array}{c}\lambda
h^{o(e)}_{n_\epsilon}(\tilde{x})\\ih^{e(o)}_{n_\epsilon-1}(\tilde{x})\end{array}\right)
\end{eqnarray*}
They have been expressed in terms of the even and odd (in $\tilde
x$) solutions $h^{e,o}_n(\tilde x)$ of the Klein-Gordon equation
$a^+a h^{e(o)}_n(\tilde x)=nh^{e(o)}_n(\tilde x)$ [the square of
(\ref{Dirac})], normalized so that $\hat a^+\,h^{e
(o)}_n=\sqrt{n+1}\;h^{o(e)}_{n+1}$, $\hat a\,
h^{e(o)}_n=\sqrt{n}\;h^{o(e)}_{n-1}$.
%\begin{eqnarray*}
%\Psi^{G,1}_{\epsilon k_y 1}(x)
%=\left(
%\begin{array}{c} h^e_{n_\epsilon-1}\left(\tilde{x}\right)\\ih^o_{n_\epsilon}\left( \tilde{x} \right)\end{array}
%\right);
%\Psi^{G,1}_{\epsilon k_y -1}(x)
%=\left(
%\begin{array}{c} h^e_{n_\epsilon}\left(\tilde{x} \right)\\ih^o_{n_\epsilon-1}\left( \tilde{x}\right)\end{array}
%\right).
%\end{eqnarray*}
%$\Psi^{G,2}$ has swapped even/odd indices respect to $\Psi^{G,1}$.
As a function of the confluent hypergeometric function
${}_1F_1(a,b,z)=1+\frac{a}{b}\frac{z}{1!}+\frac{a(a+1)}{b(b+1)}\frac{z^2}{2!}+\dots$
and $S_n=\mathrm{sign}\{\sin[\pi(n+1/2)/2]\}$, these are
\begin{eqnarray*}
h^e_n(\tilde x)&=&\sqrt{\frac{(n-1)!!}{\sqrt{\pi}n!!}}S_{n+1}\,e^{-\tilde x^2/2}\,{}_1F_1\left(-\frac{n}{2},\frac{1}{2},\tilde x^2\right),\\
h^o_n(\tilde x)&=&\sqrt{\frac{2 n!!}{\sqrt{\pi}(n-1)!!}}S_n \,\tilde
x\, e^{-\tilde
x^2/2}\,{}_1F_1\left(-\frac{n-1}{2},\frac{3}{2},\tilde x^2\right).
\end{eqnarray*}

Imposing continuity for each $\{k_y, \epsilon\}$ at the interfaces
results in the following $s-$independent transmission probability
\begin{eqnarray}
T_{k_y,\epsilon}=|t_{k_y,\epsilon}|^2=\left|\frac{2\,g^N_{n_\epsilon}}{g^R_{n_\epsilon}-ig^I_{n_\epsilon}}\right|^2,
\label{Tfull}
\end{eqnarray}
where
\begin{eqnarray}
g^R_n&=&h^{e+}_n h^{e-}_{n-1}+h^{e+}_{n-1} h^{e-}_{n}-h^{o+}_{n}
h^{o-}_{n-1}-h^{o+}_{n-1} h^{o-}_{n},\nonumber\\
g^I_n&=&h^{o+}_n h^{e-}_{n}-h^{e+}_n h^{o-}_{n}+h^{e+}_{n-1}
h^{o-}_{n-1}-h^{o+}_{n-1} h^{e-}_{n-1},\nonumber\\
g^N_n&=&h^{e+}_{n-1}h^{e+}_{n}-h^{o+}_{n-1}h^{o+}_{n}=\frac{S_{2n+1/2}}{\sqrt{\pi
n}},\nonumber
\end{eqnarray}
are expressed in terms of the wavefunctions at the boundaries
$h^{e(o)\pm}_n=h^{e(o)}_n\left(\pm \frac{L/2}{l_B}+k_yl_B\right)$.
This gives the general transmissions $T_{k_y}$ for arbitrary doping
and magnetic field.
% \footnote{Due to numerical cancelation error
%problems, this form of $T_{k_y}$ in terms of hypergeometric
%functions is not practical for the precise computations required in
%this work. Further technical developments were necessary, which will
%be expanded upon elsewhere.}.
It reproduces previously known results at $B=0$ and non-zero doping
\cite{Titov06} as well as Eq. (\ref{TRM}) for $\mu=0$.

\begin{figure}
\includegraphics[width=7.5cm]{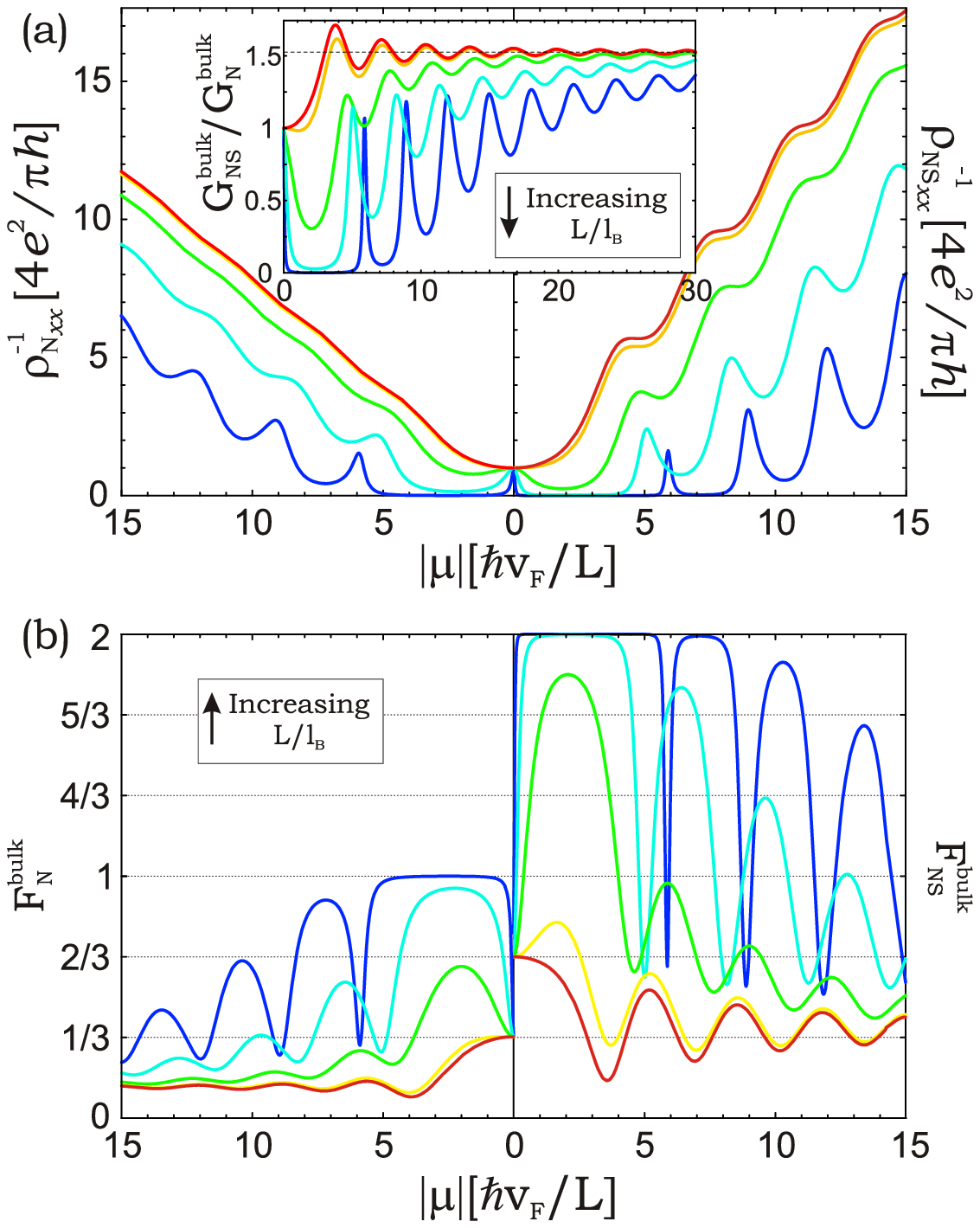}
\caption{(Color online) Inverse longitudinal resistivity in units of
$4e^2/\pi h$ (a) and bulk Fano factor (b) for the N (left) and NS
(right) junctions as a function of the (absolute value of the) Fermi
energy (in units of $\hbar v_F/L$). Different curves correspond to
different values of $L/l_B$ (where $l_B\equiv\sqrt{\hbar/eB}$),
which ranges from zero for red curve to four for the dark blue one
in steps of one. The same for the ratio
$G^{\mathrm{bulk}}_\mathrm{NS}/G^{\mathrm{bulk}}_\mathrm{N}$ in the
inset.} \label{fig:GyFvsmu}
\end{figure}

In Fig. \ref{fig:GyFvsmu}(a) we plot $\rho_{xx}^{-1}$ as a function
of the Fermi energy for increasing values of the ratio
$L/l_B\propto\sqrt{B}$. We recover the results obtained without
magnetic field, namely that $\sigma_\mathrm{N}$ and
$\sigma_\mathrm{NS}$ [where $\sigma=\rho^{-1}_{xx}(B=0)$] tend to
the known quantum-limited minimal conductivity value $4e^2/\pi h$ at
zero doping, whereas for $|\mu|L/\hbar v_F\gg 1$ the slope of the
asymptotes tends to $0.38\pi$ and $0.25\pi$ for the \textrm{NS} and
\textrm{N} junction respectively \cite{Akhmerov06}. Remarkably, as
we increase $B$, all the $\rho^{-1}_{xx}$ curves remain unchanged at
the Dirac point $\mu=0$. The Dirac-point Fano factor in Fig.
\ref{fig:GyFvsmu}(b) is also unaffected by magnetic fields, and
takes the classical diffusive value ($1/3$ for the N and $2/3$ for
the NS junction). This happens for any current cumulant, since at
$\mu=0$ the transmission given in Eq. (\ref{Tfull}) reduces to Eq.
(\ref{TRM}) independently of $B$.

However, for $\mu>0$ the resistivities and the Fano factors do
depend on the magnetic field. In particular, for $2 r_c< L$ (and
above a certain critical value of $L/l_B$) transport can take place
only at resonance with the LLs ($\mu L/\hbar v_F= \sqrt{2 n}
L/l_B$), while for other dopings $\rho^{-1}_{xx}$ is suppressed like
$e^{-(L/l_B)^2/2}$ for the N junction and as $e^{-(L/l_B)^2}$ for
the NS one. The width of the resonances at the LL energies vanishes
for $2r_c\ll L$, as we consider no disorder \cite{PGN06}.
Remarkably, $\rho_{xx}^{-1}$ at these resonances for large fields
coincides with the one at the Dirac point $4e^2/\pi h$, a
theoretical value that is usually associated strictly with zero
doping and that is interestingly at odds with some experimental
findings \cite{Netal04}. In fact it can be analytically demonstrated
that not only the conductance but the whole pseudo-diffusive
transport statistics is recovered at the resonances for high
magnetic fields. Under this perspective, the field-independent
resistivity at $\mu=0$ can be understood as due to resonant
transport through the zero-th LL that remains pinned at the Dirac
point. The field suppressed regime is apparent for small but finite
doping in Fig. \ref{fig:GyFvsmu}(a), where $\rho^{-1}_{xx}$ strongly
decreases with increasing value of $L/l_B$. Correspondingly, the
bulk Fano factor reaches the tunneling limit value ($1$ for the N
and $2$ for the NS junction) as transport gets suppressed [see Fig.
\ref{fig:GyFvsmu}(b)], in which limit the noise of the edge currents
not considered here could be visible. Increasing the Fermi energy
further one enters the regime $2r_c>L$, where $\rho^{-1}_{xx}$ is
composed of two parts. The first part is linear in $\mu L/\hbar
v_F$, in agreement with the scaling with $L$ behaviour of a
ballistic conductor subject to a magnetic field ($L$ independent
conductance). In particular, for sufficiently high dopings, all
curves in Fig. \ref{fig:GyFvsmu}(a) become parallel and tend to the
same (average) slope as the zero-field conductivity. The second
contribution to $\rho^{-1}_{xx}$ is an oscillating part, which for
$2r_c>L$ can no longer be explained by the resonance with LLs, since
in that regime the effect of the boundaries is dominating the level
structure in the central region. In fact, for $2r_c\gg L$ the
oscillations become equally spaced and are explained rather by a
Fabry-Perot type effect, connected to resonant tunneling through the
structure.

In the inset of Fig. \ref{fig:GyFvsmu}(a) the ratio
$G^{\mathrm{bulk}}_\mathrm{NS}/G^{\mathrm{bulk}}_\mathrm{N}$ is
plotted as a function of $\mu$ for the same values of $L/l_B$ as in
the main panel. At the Dirac point the ratio goes to one. At $\mu>0$
the suppressed magnetotransport manifests itself as a decaying
$G^{\mathrm{bulk}}_\mathrm{NS}/G^{\mathrm{bulk}}_\mathrm{N}\propto
e^{-(L/l_B)^2/2}$, until doping reaches the ballistic threshold and
the ratio starts growing again, finally reaching its asymptotic
value $0.38/0.25=1.52$. As explained in Ref. \cite{Akhmerov06}, this
value is expected in normal ballistic systems with Fermi wavelength
mismatch. Note again here that, for sufficiently suppressed
$G^\mathrm{bulk}$, the edge contribution \cite{Akhmerov06B}
neglected here will dominate transport.

\begin{figure}
\includegraphics[width=7.5cm]{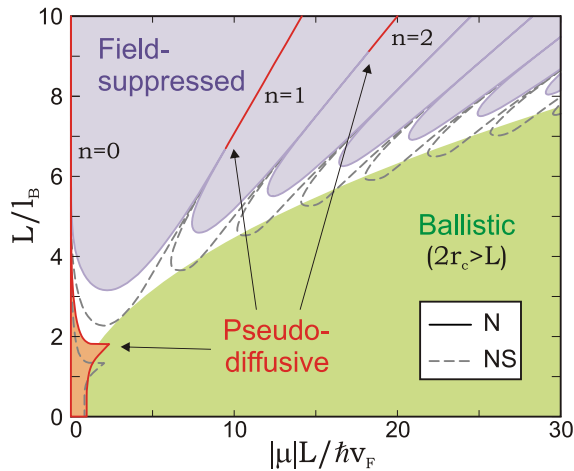}
\caption{(Color online) Phase diagram representing the crossovers
from localized, pseudo-diffusive and ballistic transport regimes in
the field-doping parameter plane. Solid (dashed) lines represent the
boundaries for a N (NS) junction. LLs are labeled by $n$.}
\label{fig:Phase-diagram}
\end{figure}

All the previous behaviours can be condensed in a quantitative way
in the phase diagram shown in Fig. \ref{fig:Phase-diagram}. It
contains three regions corresponding to the three different
transport regimes, namely: pseudo-diffusive (red), field-suppressed
(blue) and ballistic (green), in the $L/l_B$ and $\mu L/\hbar
v_F=k_F L$ parameter space. The corresponding crossover lines
between regions are solid (dashed) for the N (NS) junction (note
that the background colours correspond to the boundaries of the N
case). The boundaries for the pseudo-diffusive region have been
calculated assuming a maximum deviation of $\pm 10\%$ with respect
to the Dirac point conductivity $4e^2/\pi h$. At low fields the
width of the pseudo-diffusive window that brackets the Dirac point
is roughly field independent, whereas for $L/l_B>1.8$ for N ($1.35$
for NS) the window closes down as $\exp[-(L/l_B)^2/4]$. Physically
this means that at these higher fields the quasi-diffusive transport
regime is extremely fragile respect to doping, and an exponentially
fast crossover to the field-suppressed (localized) regime takes
place. The boundaries of the latter (blue region) were set by a
crossover criterion $\rho^{-1}_{xx}<0.1(4e^2/\pi h)$. Its spiked
shape is due to the peaked contributions to the field-suppressed
$\rho^{-1}_{xx}$ discussed in the analysis of Fig.
\ref{fig:GyFvsmu}, which are produced by resonant tunneling through
LLs. When the magnetic field is increased, the positions of these
peaks shift to higher dopings, converging on radial lines with slope
$1/\sqrt{2n}$, while their width decreases exponentially. Above a
certain value of $L/l_B$ the pseudo-diffusive regime is recovered
and the resonances are thus coloured in red. The third region
(green) is characterized by $\rho^{-1}_{xx}\propto L$ at fixed field
and doping (which would correspond to radial lines in the phase
diagram), and is therefore a ballistic transport regime. As expected
from the arguments in the introduction, the boundary of the
field-suppressed region closely follows the ballistic threshold
$2r_c=L$.
%, below
%which localization is lifted and ballistic transport sets in.
Finally, intermediate regions (white) are characterized by strongly
oscillating conductivities.

In conclusion, by computing the general transmission probabilities
through short and wide graphene junctions, we have found that the
transport properties at the Dirac point exactly match those of a
classical diffusive system even in the presence of a magnetic field,
which actually does not affect transport at all at zero doping. This
behaviour, which is associated to the existence of a zero-th LL
pinned at the Dirac point, is however found to be exponentially
fragile respect to doping for high fields. By analyzing inverse
longitudinal resistivity and higher current cumulants we have
identified and interpreted the three distinct regimes that appear at
finite magnetic fields and dopings, corresponding to
pseudo-diffusive, field-suppressed and ballistic transport, and
computed the phase diagram for the N and NS junctions in the
relevant field-doping parameter space. Transport resonances at the
LL energies are found in the field suppressed regime, with
$\rho^{-1}_{xx}$ and all higher bulk current cumulants saturating to
the pseudo-diffusive Dirac point values at high fields. The width of
these resonances decreases exponentially with magnetic field,
although broadening due to disorder in real samples is expected,
thus facilitating experimental observation. The reappearance of
pseudo-diffusive transport at finite doping could shed light on the
$1/\pi$ discrepancy between experiments and theoretical results for
the conductivity at the Dirac point.

We thank the support of G. Sch\"on in promoting this collaboration.
This work benefited from the financial support of the European
Community under the Marie Curie Research Training Networks and the
ESR program. F. G. acknowledges funding from MEC (Spain) through
grant FIS2005-05478-C02-01 and the European Union Contract 12881
(NEST).

\bibliography{GS-magnetic}
\end{document}